\documentstyle[12pt]{article}
\newcommand\beq{\begin{equation}}
\newcommand\eeq{\end{equation}}
\title{Finite size vertex corrections to the\\
three-gluon decay widths of $J/\psi$ and\\
$\Upsilon$
and a redetermination of $\alpha_s(\mu)$\\
at $\mu=m_c$ and $m_b$}
\author{H.C. Chiang\thanks{Permanent address: Institute of High
Energy Physics, Academia Sinica, Beijing, China}, J.
H\"ufner\thanks{Supported in part by the Federal Ministry of
Research and Technology under the contract number 06 HD 729}
and H.J. Pirner{\footnotesize{$^{\dag}$}}\\
Institut f\"ur Theoret. Physik der Universit\"at Heidelberg\\
and\\
Max-Planck-Institut f\"ur Kernphysik, Heidelberg, Germany}

\date{ }

\begin{document}
\setlength{\baselineskip}{24pt}
\maketitle

\begin{abstract}
\setlength{\baselineskip}{24pt}
\noindent
We calculate the corrections to the
three-gluon decay widths $\Gamma_{3g}$
of $c\bar c$ and $b \bar b$ quarkonia due to the finite extension of the
$Q\bar Q\to 3g$ vertex function. The widths computed with zero range
vertex
are reduced by a factor $\gamma$ where $\gamma=0.31$ for the $J/\psi$
and
$\gamma=0.69$ for the $\Upsilon$. These large corrections necessitate a
redetermination
of the values $\alpha_s(\mu)$ extracted from $\Gamma_{3g}$. We find
$\alpha_s(m_c)=0.28\pm.01$ and $\alpha_s(m_b)=0.20\pm.01$.
\end{abstract}

\newpage

Since the discoveries of the charmonium and
bottonium states, the physics of
heavy quark systems has been of continuous interest \cite{kwo,buc}. The
bound
state structure of heavy quarkonia becomes simpler with increasing mass,
because the constituents move more and more non-relativistically in a
static potential which is dominated for $r<0.1$~fm by the Coulomb-like
field and for $r>0.2$~fm by the confining potential.
The Schr\"odinger equation
with this potential reproduces successfully the energies of the bound
states $1s,2p,...$. The leptonic widths $\Gamma(Q\bar Q\to\mu\mu)$ of the
$^3S_1$-states, modified by radiative corrections, can also be obtained
by an expression which is proportional
to $|\psi(0)|^2$, where $\psi(\vec r)$
is the non-relativistic bound
state wave function. The value $|\psi(0)|^2$
is rather sensitive to the potential \cite{rich,kue}.
Problems exist with the
gluonic width $\Gamma(Q\bar Q\to 3g)$ and with the photo-gluonic width
$\Gamma(Q\bar Q\to\gamma gg)$. One usually calculates these
quantities in the approximation of a zero-range vertex, i.e. $\Gamma(Q
\bar Q\to 3g)\propto|\psi(0)|^2$ and analogous for $\Gamma(Q\bar Q\to
\gamma 2g)$. From a comparison with experiment one extracts a value for
the
strong interaction coupling constant $\alpha_s(\mu)$, where $\mu$ is the
scale $(\mu=m_c$ for $J/\psi$ and $\mu=m_b$ for $\Upsilon$)
\cite{kwon,kob}.
The analysis usually
proceeds in three steps. In the first step,
expressions for the ratios $R_1$
and $R_2$
\begin{eqnarray}
R_1&=&\frac{\Gamma(Q\bar Q\to\gamma gg)}{\Gamma(Q\bar Q\to
ggg)}\nonumber\\
R_2&=&\frac{\Gamma(Q\bar Q\to3g)}{\Gamma(Q\bar
Q\to\mu\mu)}\label{aa}
\end{eqnarray}
are derived in leading order $\alpha_s$.
In this order and using the zero-
range
approximation for the vertex, the decay rates are obtained from analogous
calculations for the three-photon decay of the triplet state of
positronium with the photon-electron coupling replaced by the $SU3_c$
coupling of gluons to quarks. Because of the zero-range approximation,
the rate is proportional to $|\psi(0)|^2$, the square of the
non-relativistic wavefunction, which cancels out in the ratios $R_1$
and $R_2$. In the second step,
the leading order result for $R_1$ and $R_2$
is
multiplied by the first order
$\alpha_s$ corrections, which represent a factor
of the form $(1+B(\mu)\alpha_s(\mu)/\pi)$. The coefficient $B(\mu)$
depends
both on the scale $\mu$ and the renormalization scheme.
In a third step, one usually takes into account corrections of order
$\langle\vec
q^2/m^2\rangle$ phenomenologically \cite{kwon}. These corrections may
come from the
finite extension of the decay vertex and
from relativistic corrections in the
wavefunction or from other nonperturbative physics like condensates. In
this paper we
give the first ab initio calculation of
the effect from the finite extension of
the
decay vertex, which we take into account by a calculated factor $\gamma$,
where
$\gamma=1$ represents the zero range situation.
The status of theoretical calculations
for $R_1$ and $R_2$ is given in Table 1 for the $J/\psi$ and
$\Upsilon$ together with the experimental
values for the two ratios. Values
for
$\alpha_s(\mu)$ extracted from a comparison
of the theoretical expressions
($\gamma=1$)
with the data are also shown in this table.

The following observations can be made
for the zero-range case: (i) The ratios $R_1$ and $R_2$ give
rather consistent values for $\alpha_s(\mu)$ for the same decaying system
($J/\psi$ or $\Upsilon$). (ii) Using the values
for $R_2$ which have small
error
bars, one obtains precise values for $\alpha_s(m_c)$ and $\alpha_s(m_b)$
from $J/\psi$ and $\Upsilon$ decays, respectively.
The two values are nearly equal. This result is
unexpected, because in a simple-minded QCD analysis, the scale dependence
of $\alpha_s(\mu)$ would predict
\beq
\alpha_s(m_c)\simeq\alpha_s(m_b)\frac{\ln m^2_b/\Lambda^2}
{\ln m^2_c/\Lambda^2}\simeq0.28\pm.02 \label{ab}\eeq
instead of $\alpha_s(m_c)=0.19\pm.01$ where we used
$\Lambda=200$~MeV.
This discrepancy has been observed before \cite{kwon} and
a correction factor has been parametrized but
never calculated.

We improve on the previous calculations by taking the quark-momentum
dependence (or finite size) of the transition operator $Q\bar Q\to 3g$
into account. This correction cancels in the ratio $R_1$, since the
rates in the numerator and denominator are
affected in the same way. In the
ratio $R_2$, only the three-gluon state is
modified, whereas the $s$-channel
photon annihilation still samples $|\psi(0)|^2=|\int d^3q\tilde\psi(\vec
q)|^2$, where $\tilde\psi(\vec q)$ is the
Fourier transform of $\psi(\vec r)$.
The three-gluon decay amplitude can be written as
\beq
T(k_1,k_2,k_3)=\int\frac{d^4q}{(2\pi)^4}\tilde\psi_{BS}(p,q)iM(p,q,
k_1,k_2,k_3), \label{ac}\eeq
where $k_i,i=1,2,3$ are the four momenta of the gluons, $M$ is the
expression
for the $Q\bar Q\to3g$ vertex and $\tilde\psi_{BS}(p,q)$ stands for the
Bethe-Salpeter wavefunction of the quarkonium state.
In eq.~(\ref{ac}) $p$ denotes the c.m.
momentum and $q$ the relative momentum of the two quarks.
In our calculation, we replace $\tilde\psi_{BS}$ by the non-relativistic
wavefunction $\tilde\psi(q)$ according to \cite{kueh}
\beq
\tilde\psi_{BS}(p,q)=\frac{2\pi}{i}
\delta(q^0)u(p/2+q)\tilde\psi(\vec q)\bar
v
(p/2-q). \label{ad}\eeq
The reduction of the Bethe-Salpeter wavefunction to the nonrelativistic
wavefunction
is a necessary approximation to proceed with the calculation. At short
distances,
i.e. for high momenta $|\vec p|>m_Q$,
it is expected that the relativistic
wavefunction falls off more slowly than the nonrelativistic wavefunction,
since the
relativistic transverse one-gluon
exchange kernel is less damped. Only a part
of this
effect is taken into account in the
radiative corrections \cite{mac} $\propto
0(\alpha_s)$ which include transverse gluon exchange between the heavy
$Q$ and $\bar
Q$.
The amplitude $M$ factorizes into an
expectation value for the colour part of
the gluons with fixed colors $i,j,k$ and the rest
\begin{eqnarray}
M(p,q,k_1,k_2,k_3)&=&M_{ijk}\mbox{Tr}\left(\frac{\lambda_i}{2}
\frac{\lambda_j}{2}\frac{\lambda_k}{2}\frac{1}{\sqrt3}\right)\nonumber\\
&=&\frac{1}{4\sqrt3}(d_{ijk}M^s(p,q,k_1,k_2,k_3)+if_{ijk}
M^a(p,q,k_1,k_2,k_3).
\label{ae}
\end{eqnarray}
The first part $M^s$ includes the symmetric sum of the six diagrams
arising from the interchange of the three gluons, while the antisymmetric
part $M^a$ sums the same diagrams with a negative sign for the odd
permutations. $d_{ijk}$ and $f_{ijk}$ are the structure constants of
the group $SU_3$.
For $\vec q=0$, $M^a=0$, and one finds for the three-gluon
rate the three-photon decay rate of positronium with the modification
$\alpha^3\to\frac{5}{18}\alpha^3_s$. For $\vec q\not=0$, the
contribution of $M^a$ is numerically found non-zero but negligible
in comparison to $M^s$. The vertex $M$ is
calculated from the expression
\begin{eqnarray}
&&M_{123}(p,q,k_1,k_2,k_3,\epsilon_1,\epsilon_2,\epsilon_3)=\nonumber\\
&&g^3\mbox{Tr}\left\{\not\epsilon_1\frac{1}{-\not p/2+\not q
+\not k_1-m_Q}\not\epsilon_2
\frac{1}{\not p/2+\not q-\not k_3-m_Q}\not\epsilon_3\cdot P_{s_z}
(p,q)\right\}. \label{af}\end{eqnarray}
Here, the vectors $\epsilon_i,i=1,3$
describe the gluon polarization and $P_{s_z}
(p,q)$ is the spin projection operator in the $^3S_1$ state \cite{kueh}.
We calculate the decay width $\Gamma(Q\bar Q\to3g)$ including the full
$\vec
q$-dependence in the matrix element $M$
\begin{eqnarray}
&&\Gamma(Q\bar Q\to3g)= \label{ag}\\
&&\frac{1}{3!}\sum_{\mbox{pol}}\int\frac{d^3k_1d^3k_2d^3k_3}
{8\omega_1\omega_2\omega_3(2\pi)^9}(2\pi)^4\delta^3(\Sigma
k_i)\delta(\Sigma\omega_i-2m_Q)|T(\vec k_1,\vec k_2,\vec
k_3,\vec\epsilon_1,
\vec\epsilon_2,\vec\epsilon_3)|^2,
\nonumber \\
&&T(\vec k_1,\vec k_2,\vec
k_3,\vec\epsilon_1,\vec\epsilon_2,\vec\epsilon_3)=
\int\frac{d^3q}{(2\pi)^3}M(p,\vec q,\vec k_1,\vec k_2,\vec k_3,
\vec\epsilon_1,\vec\epsilon_2,\vec\epsilon_3)\cdot\psi(\vec q).\label{ah}
\end{eqnarray}
The extensive numerical calculation to determine $\Gamma$ is performed in
the
following way. First we use the symbolic program REDUCE to evaluate the
trace in
eq.~(\ref{af}). The output of this symbolic
program is a rather long function
$M$
(50~kbytes) of $\vec q,\vec k_1,\vec k_2,\vec k_3,
\vec\epsilon_1,\vec\epsilon_2,\vec\epsilon_3$.
The $T$-matrix (\ref{ah}) is
obtained
after integrating $M\cdot\psi(\vec q)$
over $q_x,q_y,q_z$ with an adaptive
routine
which gives a precision of 5\%.
We use wavefunctions
from a numerical solution of the Schr\"odinger
equation with Richardson's potential
\cite{rich} given in ref.~\cite{kue}.
This potential goes over into the Coulomb potential for short
distances and becomes a confining linear potential at large distances.
Richardson's
potential gives a wavefunction $\psi(q)$ which falls off asymptotically as
$\alpha_s(\vec q^2)/|\vec q|^4$. The matrix element $M(q)\propto|\vec q|$
(cf. fig.~1)
together with the wavefunction leads analytically
to a very weakly diverging
integral
proportional to $\ln\ln|\vec q|_{\mbox{max}}$, which is in agreement with
our
numerical calculations. These show no sensitivity
to a variation of the cutoff
between $|\vec q|_{\mbox{max}}=3m_Q$ and $|\vec
q|_{\mbox{max}}=7m_Q$.

Finally the square of the $T$-matrix is
summed over the gluon polarization
vectors
and gluon momenta $\vec k_1=(0,0,\omega_1),\ \vec
k_2=(-\sqrt{1-x^2}\omega_3,0,-\omega_1-x\omega_3),\ \vec
k_3=(\sqrt{1-x^2}\omega_3,0,x\omega_3)$.
Energy conservation in eq.~(\ref{ag}) fixes $x$:
\beq
x=\frac{1}{2\omega_1\omega_3}[(2m_Q-\omega_1-\omega_3)^2-
\omega_1^2-\omega_2^2].
\label{ai}\eeq
The two polarization vectors of each gluon are chosen to be
$\vec\epsilon_i(1)=\vec
e_y$ and $\vec\epsilon_i(2)=\vec k_i\times\vec e_y/|\vec k_i|$. We use a
seven point
two-dimensional integration \cite{koe}
routine to do the final phase space
integration
over the Dalitz triangle given by $0\leq\omega_1\leq m,\
0\leq\omega_2\leq m$ and
$\omega_1+\omega_2\geq m$ (cf. eq.~(\ref{ag})). We estimate our total
numerical
accuracy to be 10\%. For consistency we checked our routines by setting
$\vec q=0$ in
the matrix element $M$, which leads to the zero-range approximation
$T^{(0)}$. The
corresponding decay width $\Gamma^0(Q\bar Q\to3g)$ agrees with the well-
known result
for positronium decay modified by the
gluonic couplings (eq.~(\ref{ae})) and
the strong
interaction wavefunction $\psi(r=0)$. We denote by
\beq
\gamma=\Gamma(Q\bar Q\to 3g)/\Gamma^{(0)}(Q\bar Q\to3g) \label{aj}
\eeq
the reduction factor for the three-gluon
decay due to the finite size effect in
the
matrix element $M(\vec q)$. We obtain
\beq
\gamma=\left\{ \begin{array}{lll}
0.31\pm0.03&\mbox{for}&J/\psi\\
0.69\pm0.07&\mbox{for}&\Upsilon\ .
\end{array}\right.\label{aaj}
\eeq
At first sight this correction is surprisingly
large, especially for charmonium.
In
order to demonstrate the origin of the
finite-size reduction, we plot in fig.~1
the
two factors in
the integral eq.~(\ref{ah}),
the radial Gaussian wavefunction $\vec q^2
\tilde\psi(|\vec q|)$ for $J/\psi$ and
$\Upsilon$ as a function of $|\vec q|/m_Q$ where $m_Q=m_c$ for the
$J/\psi$
and $m_Q=m_b$ for the $\Upsilon$, together with the matrix element $\bar
M^s
(\vec q,k_1,k_2,k_3,\vec\epsilon_1,\vec\epsilon_2,\vec\epsilon_3)$
which has been averaged over the directions of $\vec q$,
but is still a function of the polarization
directions. The matrix element
$\bar M^s(q)$ has its maximum at $q=0$  and falls off to a minimum at
$q/m_Q\approx1$. The exact shape depends on the gluon polarization
vectors and
the spin projection $s_z$. We have checked
the shape of the $q$-dependence
of $M^s(q)$
for various combinations of momenta $\vec
k_i$ and polarizations $\vec\epsilon_i$.
The shown case is rather typical.
It is evident from fig.~1 that $\bar M^s(q)$
cannot be well approximated by a parabolic function
\beq
\bar M^s(q)=\bar M^s(0)(1-a\ q^2/m^2_Q) \label{ak}\eeq
over the whole domain of values $q$ of interest. If we approximate
$\tilde\psi(q)$ by a Gaussian,
the form of eq.~(\ref{ak}) would lead to
a correction factor
\beq
\gamma^{est}\approx\left(1-a\frac{<v^2>}{c^2}\right), \label{al}\eeq
where $<v^2>$ is the expectation value of the velocity of the quarkonium
state
under consideration \cite{buch}.
The authors of ref.~\cite{kwon}
\underline{assume} the correction factor to be of the form eq.~(\ref{al})
and fit the coefficient $a$. They find $a$ between
2.9 and 3.5. For $a=3$, the correction amounts to $\gamma=.28$ for the
$J/\psi$
and $\gamma=.78$ for the $\Upsilon$.
These values are not too far from our
numerical values eq.~(\ref{aj}). This agreement is gratifying, since it
suggests that the finite-size correction is
probably the most important effect.

The correction factor $\gamma$ calculated in
this paper leads to new values
of $\alpha_s(m_c)$ and $\alpha_s(m_b)$ if $\gamma$ is inserted into the
theoretical expressions of $R_2$ in Table~1.
(The ratio $R_1$ is unaffected
by this correction.) The QCD corrections to the gluonic width for heavy
mesons
have been calculated in ref.~\cite{mac} for the $\Upsilon$-meson.
Special care is taken in this reference to avoid double counting of the
one gluon Coulomb exchange. The contribution to radiative corrections
due to the Coulomb exchange between $Q$ and $\bar Q$ must be dropped
(i.e. class $f$ in fig.~1 of ref.~\cite{mac}), because it is included
already in the wavefunction. Only transverse
gluons contribute to this class
of radiative corrections. Ref.~\cite{kwon} provides a practical concise
summary of all QCD corrections to the decays of quarkonia, also for
$c\bar c$-quarkonia.
The values of $\alpha_s(\mu)$ deduced
from the ratio $R_2$ are shown in Table~1. The new values of $\alpha_s$
calculated with $\gamma\not=1$
have to be compared with those from $R_1$ and are consistent.
Now the values for $\alpha_s(m_c)$ and $\alpha_s(m_b)$ are
rather different from each other,
as is expected for the running coupling constant. We can use the
values for $\alpha_s(\mu)$ to determine the QCD scale parameter
$\Lambda=\Lambda^{(4)}_{\overline{MS}}$,
which is related to $\alpha_s(\mu)$
by
\beq
\alpha_s(\mu)=\frac{12\pi}{(33-2n_f)\ln\mu^2/\Lambda^2}
\left(1-\frac{6(153-19n_f)}{(33-2n_f)^2}
\frac{\ln(\ln\mu^2/\Lambda^2)}{\ln(\mu^2/\Lambda^2)}+...\right).
\label{am}\eeq
We use $n_f=4$ and
$\mu=m_c=1.5$~GeV for $J/\psi$ and $\mu=m_b=4.9$~GeV for $\Upsilon$.
We take the values of $\alpha_s(\mu)$ with the finite size corrections
from Table~1 (for $\Upsilon$ we use an average of the values deduced
from $R_1$ and $R_2$), insert them into eq.~(\ref{am}) and solve for
$\Lambda$. The values of $\alpha_s(\mu)$ and the calculated scale
parameters $\Lambda$ are
\beq \begin{array}{ll}
\alpha_s(m_c)=0.28\pm.01& \Lambda^{(4)}_{\overline{MS}}=200\pm20\
\hbox{MeV}\\
\alpha_s(m_b)=0.195\pm.007& \Lambda^{(4)}_{\overline{MS}}=255\pm45\
\hbox{MeV}.
\end{array} \label{an}\eeq
We have also investigated the dependence on the choice of the scale $\mu$
discussed in ref.~\cite{bro} by going from $\mu=m_Q$ to $\mu=2m_Q/3$.
The radiative corrections for the strong decay widths calculated from the
expressions of ref.~\cite{kwon} change rather drastically for the two
cases. The new values of $\alpha_s(2m_Q/3)$ and the calculated values of
the scale parameter $\Lambda^{(4)}_{\overline{MS}}$ are the following
\beq \begin{array}{ll}
\alpha_s(\frac{2}{3}m_c)=0.35\pm0.01 & \Lambda^{(4)}_{\overline{MS}}=
210\pm20\\
\alpha_s(\frac{2}{3}m_b)=0.24\pm0.01 & \Lambda^{(4)}_{\overline{MS}}=
300\pm40. \end{array}\label{aan}\eeq
Eqs.~(\ref{an},\ref{aan}) summarize the main results of our paper.
The values for $\Lambda^{(4)}_{\overline{MS}}$
agree with the present world average \cite{part} of
$\Lambda^{(4)}_{\overline{MS}}=238\pm30\pm60$~MeV, where the first
error is
statistical and systematic and the second reflects the uncertainty in the
scale.

In this paper, we have shown that the inclusion of the finite size of the
$Q\bar Q\to3g$ vertex reduces significantly the $3g$ decay widths of the
$J/\psi$ and the $\Upsilon$. The
finite size modifications necessitate a redetermination
of the values for $\alpha_s(\mu)$. These new
$\alpha_s$-values are in agreement with
the QCD evolution of the coupling constant $\alpha_s(\mu)$ with the scale
$\mu$ and lead to consistent values of $\Lambda$. The corrections due to
finite size of the $3g$ vertex can be calculated without free parameter.
Because of the large size of the reduction of the 3-gluon width of
charmonium, the
question arises whether there are not other
relativistic corrections. Indeed
we
believe that the use of a nonrelativistic
wavefunction is the most crucial
approximation in our calculation. To improve on this approximation, much
more
conceptual and numerical effort would have to be spent both on the
relativistic
bound-state problem and the radiative corrections.

{\bf Acknowledgements}: We are grateful to W. Buchm\"uller,
K. K\"onigsmann, H.J. K\"uhn
and C. Wetterich for several discussions and M. Jezabek for providing us
with the program for the calculation of the quarkonium wave functions.
H.C.C. gratefully acknowledges financial support
from the Max-Planck Society
within the exchange program with the Academia Sinica.

\subsection*{Figure Caption}
\underline{Fig. 1}: The wavefunctions $q^2\tilde\psi(q)$
for the $\Upsilon_0
(m_Q=m_b$, full line) and for the $J/\psi(m_Q=m_c$,
dashed line) are given
in arbitrary units as functions of $q/m_Q$. The matrix elements $M(q)$
(eq.~(\ref{ae})) integrated over the angle of $\hat q$ are indicated for
different spin directions of the $^3S_1$-quarkonia. Squares give $M(q)$
for $J_z=-1$; crosses for $J_z=1$ and diamonds for $J_z=0$. The $z$-axis
is defined by the momentum of one fast gluon. The energies and
polarizations
$(\omega_1=\omega_2=0.97m_Q,\ \omega_3=0.06m_Q$ and
$\vec\epsilon_1=
\vec\epsilon_2=\vec\epsilon_3=(0,1,0)$) of the gluons are chosen to show
typical variations of $M(q)$ with $q/m_Q$.

\setlength{\baselineskip}{24pt}

\begin{tabular}{|c||l|l|l|}\hline
&\multicolumn{3}{c|}{$R_1=\frac{\Gamma(^3S_1\to\gamma
gg)}{\Gamma(^3S_1\to ggg)}$}\\
\cline{2-4}
& Exp & Theory & $\alpha_s(\mu)$\\
\hline
$J/\psi$ & $0.10\pm0.04$ & $\frac{16\alpha}{5\alpha_s}(1-
3.0\frac{\alpha_s}{\pi})$ &
$0.19\pm\begin{array}{l} .09\\ .05\end{array}$\\
\hline
$\Upsilon$ & $0.0274\pm0.0016$ &
$\frac{4\alpha}{5\alpha_s}(1-2.5\frac{\alpha_s}{\pi})$ &
$0.189\pm0.011$\\
\hline
\end{tabular}

\vspace{0.7cm}

\begin{tabular}{|c||l|l|l|l|l|}\hline
&\multicolumn{5}{c|}{$R_2=\frac{\Gamma(^3S_1\to
ggg)}{\Gamma(^3S_1\to\mu\mu)}$}\\
\cline{2-6}
& Exp & Theory & $\alpha_s(\mu);\gamma=1$ & $\gamma$ &
$\alpha_s(\mu)$\\
\hline
$J/\psi$ & $10.4\pm0.7$ & $\frac{5(\pi^2-
9)\alpha^3_s}{18\pi\alpha^2}(1+1.59
\alpha_s/\pi)\cdot\gamma_\psi$ & $0.190\pm.004$ & $0.31\pm0.03$
& $0.28\pm.01$\\
\hline
$\Upsilon$ &$32.5\pm0.9$ &
$\frac{10(\pi^2-9)\alpha^3_s}{9\pi\alpha^2}(1+0.43\alpha_s/\pi)
\cdot\gamma_\Upsilon$ &
$0.181\pm.02$ & $0.69\pm0.07$ & $0.20\pm.01$\\
\hline
\end{tabular}

\begin{quote}
\underline{Table 1}: Experimental values and
theoretical expressions for the
ratios $R_1$ and $R_2$ for the $J/\psi$ and $\Upsilon$ states. The
experimental
values are taken from Kwong et al. [5] for
$R_1(J/\psi)$ and from the review
talk by Kobel [7]. We used $\alpha^{-1}(m_\psi)=133.7$ and $\alpha^{-
1}(m_\mu)
=132.0$ [6].
The coupling constants $\alpha_s(\mu)$ are computed from the
experimental ratios with the help of the
theoretical expressions. Finite-range
corrections cancel in the ratio $R_1$.
In the case of $R_2$, the finite-range
corrections are contained in the factors $\gamma$. The values for
$\alpha_s(\mu)$
obtained in the zero-range approximation $\gamma=1$ are compared with
those which
include the finite-range corrections.
\end{quote}

\end{document}